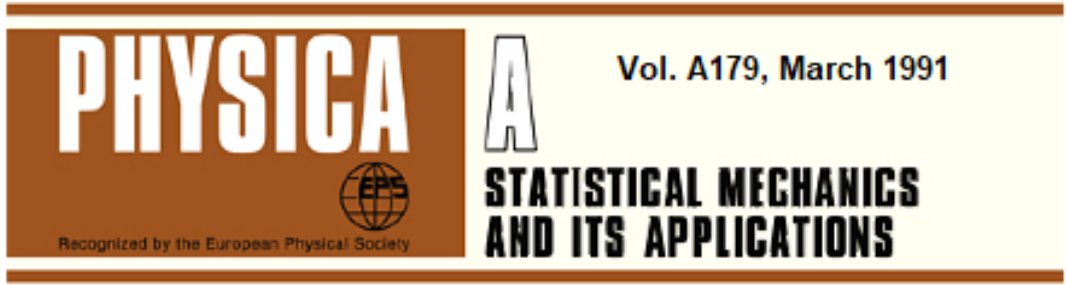

# Theory of Interfacial Tension of Partially Miscible Liquids

**M.-E. BOUDH-HIR** and **G.A. MANSOORI** *
University of Illinois at Chicago
(M/C 063) Chicago, Illinois USA 60607-7052

**Abstract**

The aim of this work is to study the problem of the existence of a fundamental relation between the interfacial tension of a system of two partially miscible liquids and the surface tensions of the pure substances. It is shown that these properties cannot be correlated from the physical point of view. However, an accurate relation between them may be developed using a mathematical artifact. In the light of this work, the basis of the empirical formula of Girifalco and Good is examined. The weakness of this formula as well as the approximation leading to it are exposed and discussed and, a new equation connecting interfacial and surface tensions is proposed.

---

(*) Corresponding author, email: *mansoori@uic.edu*

**I. Introduction**

There is a considerable interest in the understanding of inhomogeneous systems, i.e., two-phase systems, fluids in contact with a solid or subjected to an external potential, etc. This is because:

*(i)* From the fundamental point of view, it is of interest to know how to explain the phenomena related to the inhomogeneity of the system.

*(ii)* From the technical point of view, that may help to predict the properties of such systems. Indeed, the most vital fluids in our lives (water, oils, hydrocarbons, etc) exist in nature, in reserves or while in transportation in the state of confined systems. Some of these fluids are partially immiscible.

In these cases therefore, the adsorption, the interfacial tension and the surface tensions are, among others, important properties that one has to know. Unfortunately, the interparticle potentials for realistic fluids are often unknown. Hence one has to predict the behaviour of such systems using the general properties of the intermolecular potential functions (i.e., the magnitude of the potential functions, their range and the manner in which they decrease).

The study of the two-phase fluids (i.e., liquid-gas or liquid-liquid) is probably the most complex one. For this reason, few fundamental results are available in this area.[1-2] For instance, to predict the interfacial tension and the surface tensions for two-phase systems, empirical formulae[3,4] are widely used:

*(i)* Macleod's formula[3] and its various versions are often applied to the studies of liquid-vapour systems. (For more details, see for instance, reference 5).

*(ii)* In the case of two-partially miscible liquids, *a* and *b*, the formula of Girifalco and Good[4] is used. This equation is based on an analysis consisting in the assumption that the interactions between particles satisfy the so called Berthelot combining rule. According to this combining rule the potentials, assumed to be universal functions, are such that the magnitude of the interactions between two particles of different species is equal to the geometric mean of the magnitudes of the two interactions between identical particles. The interfacial tension, $\gamma$, is then written in the form:

$$\gamma = \gamma_a + \gamma_b - 2K(\gamma_a \gamma_b)^{1/2}, \tag{1}$$

where $\gamma_a$ and $\gamma_b$ are, respectively, the surface tensions of the pure liquids *a* and *b*, and $K$ is a constant depending on the nature of the system (see, for example, reference 4).

The formula of Girifalco and Good as well as the formula of Macleod, besides having very simple analytical forms, give results in good agreement with experiments for several fluids. These qualities make them widely used and their understanding is of great interest. The formula of Macleod has been derived[6,7] using the thermodynamic equation relating the surface tension, $\gamma_a$, of the system to its superficial internal energy, $u_a$,

$$u_a = S(\gamma_a - T \partial_T \gamma_a). \tag{2}$$

In this equation, $\partial_T$ denotes the differentiation with respect to the temperature. Using the statistical mechanical definition of the surface tension, it has been shown that this property is, at the first approximation, given by Macleod's formula[8]. Concerning equation (1), some questions still have to be clarified. Indeed, despite it is believed that in the liquid-vapor and liquid-liquid systems the one-particle densities have the same behavior (i.e., these functions go from their values in the first phase to their values in the second smoothly and within a thin transition region, see figure), we do not know:

   *(i)* How liquids in which the particles interact via analogous potentials can be totally or partially immiscible?

   *(ii)* How interfacial tension and surface tensions can really be correlated since they have different origins. Indeed, the surface tension arises as the consequence of the conditions (i.e., pressure and temperature) under which the system exists; whereas, the interfacial tension is due to dissimilarity of the two fluids in the system?

   *(iii)* What approximation may lead to equation (1)? Indeed, interfacial tension and surface tensions are obtained by first derivation of the free energy with respect to the surface area. It is therefore impossible to derive the third term in the right hand side of equation (1).

   *(iv)* How the interfacial tension of a liquid-liquid system decreases when the mutual miscibilities of the two liquids increase? It vanishes rigorously when the two liquids become totally miscible (i.e., at the critical temperature). In contrast, the surface tensions of the pure liquids are governed by the conditions under which the systems exist. Therefore they are not necessarily zero at the critical temperature of the system *a-b*. It follows that even if equation (1) satisfies the condition that the interfacial tension is zero at the critical temperature, the physical phenomenon is violated. Indeed, $\gamma$ may vanish at this temperature provided that three terms $\gamma_a$, $\gamma_b$ and $-2K(\gamma_a\gamma_b)^{1/2}$ compensate each other without being separately zero. This is of course not sufficient. Because, when *a* and *b* mix together, the mixture is then a homogeneous system and therefore all the contributions to the interfacial tension vanish independently. Consequently, $\gamma$ is zero without need of compensation. This condition is much more restrictive and can never be satisfied by equation (1).

   This last point implies that interfacial tension and surface tensions cannot be physically correlated and therefore, any relation between these properties is necessarily obtained by a mathematical artifact.

   The purpose of this paper is therefore to clarify these questions and, develop a new equation relating interfacial tension and surface tensions. We start by examining the case of simple liquids. Then, we consider more realistic systems in which the particles may have complex forms and interact via potentials depending not only on their separations but also on their mutual orientations. The simplifying assumptions that these interactions are additive pairwise potentials is however introduced. But this is a realistic approximation. Indeed, it has been shown that a three-particle potential can be replaced by an effective pair-potential.[9,10]

**II. Model and theoretical developments**
**A. Modeling**

Let us consider a system of particles of species *a* and *b*. The pair-potentials, $w_{aa}(i,j)$, $w_{bb}(i,j)$ and $w_{ab}(i,j)$, in this system denote, respectively, the interactions between two particles of the same species *a* or *b*, and the coupling-potential (i.e., the interaction between two particles of different species). These interactions, in the most general case, can be written as functions of the orientations, $\Omega_i$ and $\Omega_j$, of the particles *i* and *j* and the vectors $r_{ij}$ joining their centers. We have:

$$w_{\mu\nu}(i,j) = w_{\mu\nu}(r_{ij}, \Omega_i, \Omega_j), \tag{3}$$

which may be decomposed into two parts as follows:

$$w_{\mu\nu}(i,j) = v_{\mu\nu}(i,j) + u_{\mu\nu}(i,j). \tag{4}$$

Here the subscript, $\mu\nu$ stands for *aa*, *bb*, or *ab*, and $v_{\mu\nu}(i,j)$ is the repulsive part of $w_{\mu\nu}(i,j)$. It is a short-ranged function but not necessarily a hard-sphere potential. The total potential energy, *U*, of the system is:

$$U = \sum_{i<j} w_{aa}(i,j) + \sum_{i<j} w_{bb}(i,j) + \sum_{i,j} w_{ab}(i,j). \tag{5}$$

In this equation, the first two sums are extended, respectively, to the numbers $N_a$ of particles *a* and $N_b$ of particles *b*; in the third sum, *i* varies from *1* to $N_a$ and *j* is contained between *1* and $N_b$.

When two liquids are mixed together, it is well known that:

   (i)   Fluids of similar chemical nature tolerate the coexistence of each other and hence lead to homogeneous solutions as the mixture water-ethyl alcohol.
   (ii)  Fluids having very different chemical natures such as water and mercury or water and lipids are completely immiscible.

This means that the coupling-potential, (its repulsive part) plays a central role concerning the miscibility or the immiscibility of the two fluids. Indeed, a strong repulsion between particles *a-b*, compared to the repulsion between particles *a-a* or *b-b*, tends to create vacancies between particles of different species. This free space is filled by particles *a* and/or *b* in such a way that each molecule tends to have a maximum number of neighbors identical to it with the consequence that the system splits into two phases. The first, *A*, is a saturated solution of *b* in *a* and the other, *B*, of *a* in *b*. In the limiting case where the difference between the interactions is very important, the two liquids become completely immiscible. In contrast, they are completely miscible when these interactions are comparable. Between the two extreme cases of totally miscible and immiscible, there exist all the intermediate situations (i.e., fluids partially miscible with compositions varying between zero and one). For systems diluted enough to have the average particle separations larger than the range of the repulsive potential parts, this difference between the particle-particle interactions' behavior does not have any consequence concerning the immiscibility of the two fluids. The increasing in densities makes the average distances between particles smaller and smaller and, the separation of the system into two phases follows.

From this model, the existence of a critical temperature beyond which the liquids are completely miscible is easily understood. The increase in temperature leads to the fact that the thermal diffusion effect becomes more important than the repulsive force between particles. In mathematical terms, the Boltzmann factors associated to the repulsive potential parts tend to the same limit and therefore the formation of a single-phase system follows.

Concerning the attractive potential part, one cannot expect that it may, in any way, generate a phase separation. Consequently, this phenomenon is due to the repulsive part of the coupling-potential and therefore, very dissimilar interaction potentials are necessary to split the system into two phases. This conclusion is supported by the explicit calculations made for hard sphere systems.[11-14] It is true that some differences between idealized and realistic systems exist but, the observed physical phenomena should remain essentially the same. In the previous work, binary mixtures of hard spheres (*a* and *b*) characterized by their hard core diameters $\sigma_{aa}$, $\sigma_{bb}$ and $\sigma_{ab}$ satisfying the condition:

$$\sigma_{ab} = (1 + \delta)(\sigma_{ab} + \sigma_{bb})/2, \tag{6}$$

have been studied. It has been shown that:

*(i)* There is no phase separation when $\delta$ is equal to zero (mixtures of additive hardcore diameters).[11]

*(ii)* For negative values of $\delta$, the systems have compound-forming tendency.[12,14]

*(iii)* For $\delta$ greater than zero (of the order of $10^{-2}$), the mixture leads to two-phase systems at densities depending on this parameter.[13] For very small $\delta$ values, the transition density is in the solid region and the miscible state is frozen before the two-phase transition can occur.

**B. Theoretical developments**

The interfacial tension, $\gamma$, is defined as:

$$\gamma = (\partial_S F)_{V,T,\mu}$$

$$= 1/2 \int d1 d2\, \rho_{aa}(1,2) \left((\partial_S w_{aa}(1,2))\right)_V + 1/2 \int d1 d2\, \rho_{bb}(1,2) \left((\partial_S w_{bb}(1,2))\right)_V$$

$$+ \int d1 d2\, \rho_{ab}(1,2) \left((\partial_S w_{ab}(1,2))\right)_V$$

$$= \gamma_a^* + \gamma_b^* + \gamma_{ab}^*. \tag{7}$$

Here, $F$ stands for the free energy of the system under consideration, $V$, $T$ and $\mu$ are, respectively, the volume, temperature and chemical potential characterizing this system and $\rho_{aa}(1,2)$, $\rho_{bb}(1,2)$ and $\rho_{ab}(1,2)$ denote the two-particle densities.

It should be noted that:

*(i)* The first two terms, $\gamma_a^*$ and $\gamma_b^*$, in this equation have the same expression as the pure substances $a$ and $b$ surface tensions, respectively. This analogy is however purely formal. Indeed, the two-particle densities, $\rho_{aa}(1,2)$ and $\rho_{bb}(1,2)$ take values in the partially miscible fluid system $a$-$b$ different from those taken by their analogues, $\rho^0_{aa}(1,2)$ and $\rho^0_{bb}(1,2)$, in the one-component systems $a$ and $b$.

*(ii)* The term $\gamma_{ab}^*$ cannot be compared to a pair-potential contribution to the surface tension of a one-component system because the two particles *1* and *2* are of different species. Because none of these particles can be considered fixed in the space, $\gamma_{ab}^*$ cannot be compared to the contribution of an external potential to the surface tension of a one-component system.

*(iii)* For fluids of similar chemical nature or existing at a temperature higher than the critical temperature of the mixture, the mixture is a homogeneous system. Therefore, the three terms, $\gamma_a^*$, $\gamma_b^*$ and $\gamma_{ab}^*$, independently become zero. Thus the interfacial tension vanishes as it should be without need of compensation and whatever the values of the surface tensions, $\gamma_a$ and $\gamma_b$, would be when the temperature of the pure liquids is equal to the critical temperature of their mixture.

Nevertheless, since for both of the two-phase systems (liquid-vapor and liquid-liquid) the transition region is a thin layer and the one-particle densities have the same behavior (see figure), $\gamma_a^*$ and $\gamma_b^*$ may be accurately approximated starting with the surface tensions, $\gamma_a$ and $\gamma_b$. We have:

$$\gamma_a^* = 1/2 \int d1d2\, \rho^0_{aa}(1,2)\, (\partial_S w_{aa}(1,2))_V \{\rho_{aa}(1,2)/\rho^0_{aa}(1,2)\}$$

$$= 1/2 \int d1d2\, \rho^0_{aa}(1,2)\, (\partial_S w_{aa}(1,2))_V\, \Phi_{aa}(1,2), \qquad (8a)$$

$$\gamma_b^* = 1/2 \int d1d2\, \rho^0_{bb}(1,2)\, (\partial_S w_{bb}(1,2))_V \{\rho_{bb}(1,2)/\rho^0_{bb}(1,2)\}$$

$$= 1/2 \int d1d2\, \rho^0_{bb}(1,2)\, (\partial_S w_{bb}(1,2))_V\, \Phi_{bb}(1,2), \qquad (8b)$$

The term $\gamma_{ab}^*$ can be re-expressed as:

$$\gamma_{ab}^{*2} = \int d1d2\, \rho^0_{aa}(1,2)\, (\partial_S w_{aa}(1,2))_V \{\rho_{ab}(1,2)\,(\partial_S w_{ab}(1,2))_V / \rho^0_{aa}(1,2)\,(\partial_S w_{aa}(1,2))_V\}$$

$$\int d1d2\, \rho^0_{bb}(1,2)\, (\partial_S w_{bb}(1,2))_V \{ \rho_{ab}(1,2)\, (\partial_S w_{ab}(1,2))_V / \rho^0_{bb}(1,2)\, (\partial_S w_{bb}(1,2))_V \}$$

$$= \int d1d2\, \rho^0_{aa}(1,2)\, (\partial_S w_{aa}(1,2))_V \Phi_{a/b}(1,2) \int d1d2\, \rho^0_{bb}(1,2)\, (\partial_S w_{bb}(1,2))_V \Phi_{b/a}(1,2). \qquad (9)$$

It is clear that the following approximations:

$$\Phi_{aa}(1,2) \approx 1, \qquad (10a)$$

$$\Phi_{bb}(1,2) \approx 1, \qquad (10b)$$

$$\Phi_{a/b}(1,2)\, \Phi_{b/a}(1,2) \approx K^2, \qquad (10c)$$

lead to equation (1) since it is evident that $\gamma^*_{ab}$ is negative. Indeed, this quantity represents the energy required to separate the two liquids $a$ and $b$ without modifying the manner in which their particles are distributed. However, it should be mentioned that since the pure substances $a$ and $b$ and the system liquid-liquid resulting from the mixture of these two liquids do not have the same critical temperature, approximation (10a),(10b) and (10c) may lead to an interfacial tension for the system $a$-$b$ which does not vanish at its critical temperature and beyond this limit. In this case, the condition that the surface tension should vanish at the critical temperature is neither satisfied from the mathematical point of view (i.e., $\gamma$ is not zero) nor from the physical point of view (i.e., the three components $\gamma^*_a$, $\gamma^*_b$ and $\gamma^*_{ab}$ are not simultaneously zero). This is the due of the fact that the interfacial tension, $\gamma$, is expanded in terms of surface tensions, $\gamma_a$ and $\gamma_b$, which are not intrinsically correlated to $\gamma$. Thus the relation given by equation (1) is obtained through a mathematical artifact.

Here we approximate the interfacial tension given by equation (7), (8a), (8b) and (9) in such a way to remove the shortcoming discussed in the above paragraph. Before examining the more general case, we consider systems in which the pair-potentials $v_{aa}(i,j)$, $v_{bb}(i,j)$ and $v_{ab}(i,j)$ conform to the same analytical forms and depend only on $r_{ij}/\sigma_{aa}$, $r_{ij}/\sigma_{bb}$ and $r_{ij}/\sigma_{ab}$, respectively. While such systems are appropriate models for simple fluids, their application is extended to polar mixtures.[15,16] Their study allows us to gain some insight as how to implement the treatment of the general case. For these kind of potentials, it has been shown that the soft repulsive interaction of parameter $\sigma_{\mu\nu}$ may be replaced by a hard sphere potential whose parameter $\sigma^*_{\mu\nu}$ is larger than $\sigma_{\mu\nu}$. By choosing the range-parameters, $\sigma_{aa}$, $\sigma_{bb}$ and $\sigma_{ab}$ satisfying condition (6), this joins the case treated in reference 13.

**III Approximation**
**A. Simple fluid case**

The differentiation with respect to the surface area in equations (7), (8a), (8b) and (9)

may be performed using the transformation:[17]

$$r = (x, y, z)$$
$$= (S^{1/2}x', S^{1/2}y', Vz'/S), \tag{11}$$

which leads to:

$$(\partial_S w_{\mu\nu}(i,j))_V = ((r_{ij}^2 - 3z_{ij}^2)/2Sr_{ij}) \partial_{r_{ij}} w_{\mu\nu}(r_{ij}). \tag{12}$$

Due to the equivalence of the three directions, $x$, $y$, and $z$, in the phases $A$ and $B$, the only contribution to the interfacial tension comes from the transition region.

Taking into account the fact that:

*(i)* The profile densities are assumed to be varying smoothly within a thin region (see figure). Therefore, in the transition region, these functions may be approximated by:

$$\rho_v(i) \approx (\Delta\rho_v/\Delta\rho_v^0) \rho_v^0(i), \tag{13}$$

*(ii)* When an inhomogeneity arises in the system, the one-particle densities are changed. Then the correlation functions will be modified as a consequence of the change in the one-particle densities. Therefore, the change in the correlation functions can be considered of second order. It follows that $\Phi_{aa}(1,2)$ and $\Phi_{bb}(1,2)$ in equations (8a) and (8b) may be, at the first approximation, replaced by:

$$\Phi_{aa}(1,2) \approx (\rho_{a/A} - \rho_{a/B})^2(\rho_a^l - \rho_a^v)^{-2}$$
$$= (\Delta\rho_a/\Delta\rho_a^0)^2, \tag{14a}$$

$$\Phi_{bb}(1,2) \approx (\rho_{b/B} - \rho_{b/A})^2(\rho_b^l - \rho_b^v)^{-2}$$
$$= (\Delta\rho_b/\Delta\rho_b^0)^2, \tag{14b}$$

where $\rho_{a/A}$ and $\rho_{a/B}$ denote the values of the profile densities of the liquid $a$ in the phases $A$ and $B$, respectively; whereas $\rho_a^l$ and $\rho_a^v$ stand for the values of the profile densities of the pure fluid $a$ in the phases liquid and vapor, respectively. The quantities related to the liquid $b$ are defined by the same way.

Using equations (8a), (8b), (14a) and (14b), $\gamma_a^*$ and $\gamma_b^*$ take the forms:

$$\gamma_a^* = (\Delta\rho_a/\Delta\rho_a^0)^2 \gamma_a$$
$$= \alpha^2 \gamma_a, \tag{15a}$$

$$\gamma_b^* = (\Delta\rho_b/\Delta\rho_b^0)^2 \gamma_b$$

$$= \beta^2 \gamma_b; \tag{15b}$$

Concerning the function $\Phi_{a/b}(1,2)$, using equations (9) and (12) it may be written:

$$\Phi_{a/b}(1,2) = \left(\rho_{ab}(1,2)\, \partial_{r_{12}} w_{ab}(1,2)\right) / \left(\rho_{aa}(1,2)\, \partial_{r_{12}} w_{aa}(1,2)\right). \tag{16}$$

Using equation (13) and averaging over $z_1$ and $z_2$, equation (16) takes the form:

$$\Phi_{a/b}(1,2) = (\sigma_{aa}/\sigma_{ab})\,(\Delta\rho_a/\Delta\rho_a^0)\,(\Delta\rho_b/\Delta\rho_b^0)\,\{\,[1/2(\rho_b^l + \rho_b^v)\,g_{ab}(r_{12}^{ab})]\,\partial_{r_{12}^{ab}} w_{ab}(r_{12}^{ab})/$$

$$[1/2(\rho_a^l + \rho_a^v)\,g_{aa}(r_{12}^{aa})]\,\partial_{r_{12}^{aa}} w_{aa}(r_{12}^{aa})\}\,. \tag{17}$$

The density of the fluid $b$ is now assumed to be equal to $1/2\,(\rho_b^l + \rho_b^v)$, $1/2\,(\rho_b^l + \rho_b^v)\,g_{ab}(r_{12}^{ab})$ is the probability density to find one particle of species $b$ at the distance $r_{12}^{ab}$ from a given particle of species $a$ where $r_{12}^{\mu\nu}$ is the distance between the two particles $1$ and $2$ of species $\mu$ and $\nu$ expressed in unit $\sigma_{\mu\nu}$. We have:

$$r_{12}^{\mu\nu} = r_{12}/\sigma_{\mu\nu}. \tag{18}$$

Introducing the approximations:

$$\{\,1/2[(\rho_b^l + \rho_b^v)\,g_{ab}(r_{12}^{ab})]/1/2[(\rho_a^l + \rho_a^v)\,g_{aa}(r_{12}^{aa})]\} \approx \{(\rho_b^l + \rho_b^v)/(\rho_a^l + \rho_a^v)\}\,(\sigma_{aa}/\sigma_{ab})^3, \tag{19a}$$

$$\partial_{r_{12}^{ab}} w_{ab}(r_{12}^{ab}) / \partial_{r_{12}^{aa}} w_{aa}(r_{12}^{aa}) \approx 1, \tag{19b}$$

$\Phi_{a/b}(1,2)$ in equation (16) becomes:

$$\Phi_{a/b}(1,2) \approx (\Delta\rho_a/\Delta\rho_a^0)\,(\Delta\rho_b/\Delta\rho_b^0)\,[(\rho_b^l + \rho_b^v)/(\rho_a^l + \rho_a^v)]\,(\sigma_{aa}/\sigma_{ab})^4. \tag{20}$$

By symmetry, we can write for $\Phi_{b/a}(1,2)$:

$$\Phi_{b/a}(1,2) \approx (\Delta\rho_a/\Delta\rho_a^0)\,(\Delta\rho_b/\Delta\rho_b^0)\,[(\rho_a^l + \rho_a^v)/(\rho_b^l + \rho_b^v)]\,(\sigma_{bb}/\sigma_{ab})^4. \tag{21}$$

Equations (20) and (21) together with (9) lead to:

$$\gamma_{ab}^* \approx -2\,(\Delta\rho_a/\Delta\rho_a^0)\,(\Delta\rho_b/\Delta\rho_b^0)\,(\sigma_{aa}\sigma_{bb}/\sigma_{ab}^2)^2\,(\gamma_a\gamma_b)^{1/2}$$

$$= -2 \alpha \beta K (\gamma_a \gamma_b)^{1/2}. \tag{22}$$

Using equations (7), (15a), (15b) and (22), the interfacial tension for a system of two partially immiscible liquids takes the form:

$$\gamma = (\Delta\rho_d/\Delta\rho^0_a)^2 \gamma_a + (\Delta\rho_b/\Delta\rho^0_b)^2 \gamma_b - 2(\sigma_{aa}\sigma_{bb}/\sigma_{ab}^2)^2 (\Delta\rho_a \Delta\rho_b/(\Delta\rho^0_a \Delta\rho^0_b)) (\gamma_a \gamma_b)^{1/2}$$

$$\gamma = \alpha^2 \gamma_a + \beta^2 \gamma_b - 2\alpha\beta K (\gamma_a \gamma_b)^{1/2}. \tag{23}$$

It should be noted that:
  *(i)* Two differences exit between this equation and that proposed by Girifalco and Good:
  First, in equation (23), $\gamma_a^*$ and $\gamma_b^*$ are associated with the coefficients $(\Delta\rho_d/\Delta\rho^0_a)^2$ and $(\Delta\rho_b/\Delta\rho^0_b)^2$, respectively. These two coefficients replaced by *1* in Girifalco and Good equation.
  Second, in equation (23), the term $(\gamma_a \gamma_b)^{1/2}$ is associated with the coefficient $-2(\sigma_{aa}\sigma_{bb}/\sigma_{ab}^2)^2 (\Delta\rho_a \Delta\rho_b/(\Delta\rho^0_a \Delta\rho^0_b))$; whereas it is associated with the coefficient $-2(\sigma_{aa}\sigma_{bb}/\sigma_{ab}^2)$ in the equation of Girifalco and Good.
  *(ii)* Besides the arising in equation (23) of the terms $\Delta\rho_a$ and $\Delta\rho_b$ which guarantee that the three components $\gamma_a^*$, $\gamma_b^*$ and $\gamma_{ab}^*$ go to zero when the temperature of the system tends to its critical value, the approaches leading to these two equations are completely different.
  *(iii)* The factor $\delta$ in equation (6) is very small. Few per cents are enough to create vacancies between particles *a* and *b* leading to the split of the system into two phases. It follows that:
  First, $\delta$ plays an indirect role in the arising of the interfacial phenomena, i.e., due to to the split of the system into two phases, the particles will not be distributed uniformly and the interfacial tension arises.
  Second, in equation, (23) $\sigma_{ab}$ can be replaced by $(\sigma_{aa} + \sigma_{bb})/2$. The direct (macroscopic) effect of $\delta$ can be neglected.
  *(iv)* If the liquid *b* in the binary mixture is now replaced by a system of *n* liquids which mix together but not miscible with the liquid *a*, equation (23) becomes:

$$\gamma = (\Delta\rho_d/\Delta\rho^0_a)^2 \gamma_a + \sum_{1 \leq i \leq n} \left\{ (\Delta\rho_i/\Delta\rho^0_i)^2 \gamma_i - 2(\sigma_{aa}\sigma_{ii}/\sigma_{ai}^2)^{1/2} (\Delta\rho_a \Delta\rho_i/(\Delta\rho^0_a \Delta\rho^0_i)) (\gamma_a \gamma_i)^{1/2} \right\}. \tag{24}$$

Because of the approximation used, it is expected that the less miscible the liquids are the better this approximation becomes.

**B. General case**
  We now turn back to the more general case of molecular fluids. The particle-particle potential, *w(i,j)*, has the form given by equation (3). It follows that its derivative with respect to the surface area will take a form a little different from equation (12):

$$(\partial_S w_{\mu\nu}(i,j))_V = (1/2S) r_{ij}^* \cdot \partial_{r_{ij}} w_{\mu\nu}(i,j), \tag{25}$$

in which the vector $r_{ij}^*$ and the operator $r_{ij}^* \cdot \partial_{r_{ij}}$ are defined respectively by:

$$r_{ij}^* = (x_{ij}, y_{ij}, -2z_{ij}), \quad (26a)$$

$$r_{ij}^* \cdot \partial_{r_{ij}} = (x_{ij} \partial_{x_{ij}}, y_{ij} \partial_{y_{ij}}, -2z_{ij} \partial_{z_{ij}}). \tag{26b}$$

For simplify the estimation of the different terms $\Phi_{aa}(1,2)$, $\Phi_{bb}(1,2)$, $\Phi_{a/b}(1,2)$ and $\Phi_{b/a}(1,2)$, it is convenient to write the molecular pair-potential as a sum of simple-fluid potential terms. Such a decomposition can be obtained by introducing the site-site interaction model. The molecular potential becomes:

$$w(i,j) = \sum_{m,n} v(i_m, j_n), \tag{27}$$

where the site-site interaction, $v(i_m, j_n)$, depends only on the distance between the sites under consideration. It can be written in the form:

$$v(i_m, j_n) = v(r_{i_m j_n} / \sigma_{mn}), \tag{28}$$

$\sigma_{mn}$ is the hard core parameter associate to these sites. Two particles of different species should have at least one couple of sites whose pair-potentials' hard core parameters, $\sigma_{aa}$, $\sigma_{bb}$ and $\sigma_{ab}$, satisfy the condition (6). The decomposition given by equation (27) is:

   *(i)* attractive. Because intuitively, it is natural to write that the interactions between two molecules are equal to the sum of the interactions between their different particles.
   *(ii)* exact. Because the pair-potential can be decomposed in any way, the only requirement is that the whole potential should be realistic.

The use of the site-site interaction model is supported by the good agreement with the results obtained from the molecular dynamics and the experiments.[18] The theory related to this model has been recently reformulated and the results obtained are in good agreement with the numerical simulations in both cases of low and high temperatures.[19]

Using the same arguments as in the case of simple fluids, equations (14a) and (14b) and, therefore equations (15a) and (15b), can be extended to the general case. Concerning the approximation of the functions $\Phi_{a/b}(1,2)$ and $\Phi_{b/a}(1,2)$ in equation (9), results obtained for pure molecular fluids[19,20] can easily be extended to their mixtures. The molecular pair-distribution functions in the mixture *a-b* can be written as:

$$g_{\mu\nu}(1,2) = \Pi_{m^3 1} \Pi_{n^3 Á} \, g^0(1_m, 2_n) \, \Psi_{\mu\nu}(1,2), \tag{29}$$

where the products over $m$ and $n$ are extended to all the sites of the particles $1$ of species $\mu$ and $2$ of species $\nu$, $g_{\mu\nu}(1,2)$ stands for the molecular pair-distribution function, $g^0(1_m, 2_n)$ are the pair-distribution functions in a mixture of simple liquids whose particles are identical to the sites of the particles $1$ and $2$, and $\Psi_{\mu\nu}(1,2)$ is a function converging rapidly to unity. An estimation for the functions $\Phi_{a/b}(1,2)$ and $\Phi_{b/a}(1,2)$ can be obtained by: *(i)* neglecting the functions, $\Psi_{\mu\nu}(1,2)$ and, *(ii)* for a given molecule, each site is replaced by a uniform distribution of sites over a sphere centered on one of the molecule sites. In the case where the liquid $a$ has spherical molecules and $b$ has linear molecules consisting of $k$ sites of equal hard core diameter, $\sigma_{ss}$, we can write:

$$\Phi_{a/b}(1,2) \approx (\Delta\rho_a/\Delta\rho_a^0)(\Delta\rho_b/\Delta\rho_b^0)[(\rho_b^l + \rho_b^v)/(\rho_a^l + \rho_a^v)](\sigma_{aa}/\sigma_{ab})^3(\sigma_{aa}/\sigma_{as})(k+2)/2. \tag{30}$$

$$\Phi_{b/a}(1,2) \approx (\Delta\rho_a/\Delta\rho_a^0)(\Delta\rho_b/\Delta\rho_b^0)[(\rho_a^l + \rho_a^v)/(\rho_b^l + \rho_b^v)](\sigma_{bb}/\sigma_{ab})^3(\sigma_{ss}/\sigma_{as})$$

$$\{(k+2)/[4[1+(k+2)^2/4]]\}, \tag{31}$$

where $\sigma_{bb}$ is the parameter associated with the spherical molecule equivalent to those of the liquid $b$ and $\sigma_{as}$ characterizes the interaction site-molecule $a$. Equations (30) and (31) together with (9) lead to:

$$\gamma^*_{ab} \approx -2 (\Delta\rho_a/\Delta\rho_a^0)(\Delta\rho_b/\Delta\rho_b^0)(\sigma_{aa}\sigma_{bb}/\sigma_{ab}^2)^{3/2}(\sigma_{aa}\sigma_{ss}/\sigma_{as}^2)^{1/2}(2(1+4/(k+2)^2))^{-1/2}(\gamma_a\gamma_b)^{1/2}$$

$$= -2 \, \alpha\beta K (\gamma_a\gamma_b)^{1/2}. \tag{32}$$

The interfacial tension has the same form as in equation (23) i.e.,

$$\gamma = \alpha^2 \gamma_a + \beta^2 \gamma_b - 2\alpha\beta K (\gamma_a\gamma_b)^{1/2}, \tag{33}$$

where the coefficient $K$ is now different that obtained for simple liquids and given by equation (22).

## IV. Concluding remarks

The purpose of this paper was to examine the problem of the existence of a fundamental relation between the interfacial tension of a system of two partially miscible liquids and their surface tensions. It has been seen that these properties cannot be physically correlated. However, because of the fact that in the two-phase systems: liquid-liquid $a$-$b$ and the one-component $a$ and $b$ liquid-vapor, the transition regions are thin layers and the one-particle densities have the same behavior, the expression of the interfacial tension can be mathematically approximated in such a way to express this property in term of $\gamma_a$ and $\gamma_b$. The equation proposed differs from the Girifalco and Good equation from two points of view:

   *(i)* The coefficients associated to the terms $\gamma_a$, $\gamma_b$ and $(\gamma_a \gamma_b)^{1/2}$ in the expression of $\gamma$ are different from the corresponding coefficients in equation (1).

   *(ii)* The three contributions to the interfacial tension vanish independently when the temperature of the mixture reaches its critical value. Consequently $\gamma$ becomes zero, as it should be, and without need of compensation between its different terms.
The accuracy of this equation is examined in the second part of this work.

## Acknowledgment

This research is supported by the Chemical Sciences Division Office of Basic Energy of the U.S. Department of Energy, GrantDE-FG84ER13229.